\documentclass[english,aps,prl,reprint,superscriptaddress]{revtex4-1}
\usepackage[T1]{fontenc}
\usepackage[utf8]{inputenc}
\usepackage{xcolor}
\usepackage{units}
\usepackage{graphicx}
\usepackage{hyperref}
\usepackage{babel}

\begin{document}

\title{Memory in 3D Cyclically Driven Granular Material}

\author{Zackery A. Benson}
\affiliation{Institute for Physical Science and Technology, College Park, Maryland, USA}
\affiliation{Department of Physics, University of Maryland, College Park, Maryland, USA}

\author{Anton Peshkov}
\affiliation{Department of Physics, University of Rochester, Rochester, New York, USA}

\author{Derek C. Richardson}
\affiliation{Department of Astronomy, University of Maryland, College Park, Maryland, USA}

\author{Wolfgang Losert}
\affiliation{Institute for Physical Science and Technology, College Park, Maryland, USA}
\affiliation{Department of Physics, University of Maryland, College Park, Maryland, USA}

\begin{abstract}
We perform experimental and numerical studies of a granular system under cyclic-compression to investigate reversibility and memory effects. We focus on the quasi-static forcing of dense systems, which is most relevant to a wide range of geophysical, industrial, and astrophysical problems. We find that soft-sphere simulations with proper stiffness and friction quantitatively reproduce both the translational and rotational displacements of the grains. We then utilize these simulations to demonstrate that such systems are capable of storing the history of previous compressions. While both mean translational and rotational displacements encode such memory, the response is fundamentally different for translations compared to rotations. For translational displacements, this memory of prior forcing depends on the coefficient of static inter-particle friction, but rotational memory is not altered by the level of friction.  
\end{abstract}
\maketitle

\section{Introduction}

The study of memory in materials is an extensive field of research with implications for our understanding of biology, condensed matter physics, and granular materials\cite{Axmacher:2006}\cite{Planes:2009}\cite{Walker:2015}\cite{Plati:2019}. Memory in materials corresponds to storing information in a material “state”.  A familiar example of a state is the local direction of magnetization, which enables magnetic information storage\cite{Chikazumi:1997}. Retrieving this memory requires knowing the material state used to store the information, e.g., that hard drives store information in magnetized regions of a certain size.  

Granular materials can also store information about their past:  it is possible to discern the direction\cite{Toiya:2004} of prior rearrangements or the amplitude of prior shear\cite{Paulsen:2014}. But the material “state” that stores this information is not yet well understood. It has been shown that systems with identical density and pressure, but different preparation history, would diverge in their future evolution\cite{Knight:1995}. Therefore, it is highly probable that, unlike in the case of magnetization, memory in granular systems is not stored in macroscopic quantities such the density or pressure, but in a complex state space involving particle positions, velocities, and contact networks\cite{Adhikari:2018}\cite{Bandi:2018}.

Previous simulations\cite{Adhikari:2018}\cite{Bandi:2018} and experiments\cite{Paulsen:2014}\cite{Mukherji:2019} have demonstrated that it is possible to retrieve granular memory by measuring particle displacements in response to periodic driving. The memory is extracted by conducting a  sweep of perturbation amplitudes, which measures the mean squared displacement (MSD) of the grains\cite{Adhikari:2018}\cite{Fiocco:2018}. Specifically, memory of the prior driving emerges when the MSD transitions from being reversible to irreversible. The majority of memory studies on granular systems have been focused on memory encoded in the linear displacements of the grains\cite{Paulsen:2014}\cite{Adhikari:2018}, with only a few works studying frictional dissipation\cite{Bandi:2018}\cite{Richard:2005}. Since friction is present in most real-world materials, it is crucial to understand how friction affects memory formation in granular materials. Further, since friction drives rotations, it is important to discern whether memory can be encoded in individual rotations of particles.

In this paper, we study memory formation in a dense 3D packing of athermal, frictional grains subject to cyclic compression. Almost all granular matter is subject to cyclic forcing in geological, astrophysical, or engineering contexts. It has been shown that cyclically driven assemblies of spheres exhibit either reversible or irreversible motion depending on the perturbation applied\cite{Walker:2015}\cite{Pine:2005}\cite{Peshkov:2019}\cite{Royer:2015}\cite{Slotterback:2012}. Consequently, these types of assemblies can act as a model system for exploring the formation and origin of memory in granular materials. We expand our study to measure all aspects that characterize the system, including grain rotations and translations, and their role in encoding memory. We do this numerically using soft-sphere discrete element method (DEM) simulations that we calibrate using experimental data. Finally, we demonstrate that memory does form in our assembly of spheres by measuring the MSD of our grains between cycles, and we probe the effect frictional contacts have on the evolution of our system. 

\begin{figure*}
\includegraphics[width=1.8\columnwidth]{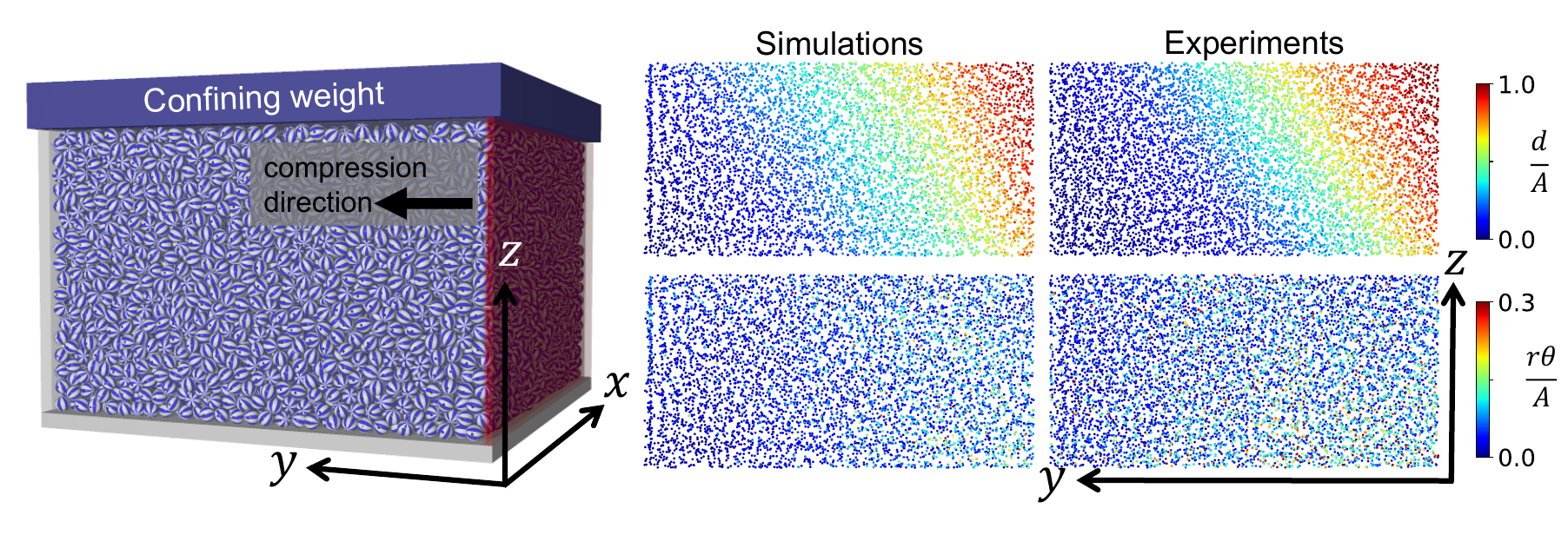}
\caption{\label{fig:one}Spatial distribution of displacements (top row) and rotations (bottom row) for the simulations (left) and experiments (right) at full compression. Color indicates magnitude of motion, with red and blue being large and small displacements, respectively; both the simulation and the experiment are on the same scale. The compression amplitude is 1.0\% of the container size. Schematic: snapshot of the simulation setup. The pattern on the beads is a visual help to observe the orientation. Compression wall is in red and compresses along the y-axis.}
\end{figure*}

\section{Methods}

Experiments - The experimental system consists of a monodispersed mixture of 20,000 acrylic, spherical grains with diameters of 0.5 cm, which possess a cylindrical cavity across their center that is used to track the rotations of the particles. The grains reside in a square-based container with a side length of 15 cm up to a height of approximately 10 cm and are immersed in an index-matching solution (Triton X-100, n = 1.49), which allows us to track the translations of the particles as well as their rotations. A free weight of 1 kg is placed on top to achieve a constant external pressure. The system is sheared by compressing a lateral wall horizontally along a single axis by an amplitude A and a whole cycle is completed when the wall compresses and then fully decompresses the system. We consider the response of the system to repeated compression cycles. Additional information on the experimental setup can be found in references 
\cite{Peshkov:2019}\cite{HARRINGTON:2014}.

Simulations - Our numerical model consists of soft-sphere DEM simulations using an in-house software package\cite{Schwartz:2012}. A linear spring is used to calculate the forces between grains, with a spring constant chosen to maintain much less than 1\% overlap between grains. A friction model is included consisting of static, rolling, and twisting friction (see \cite{Schwartz:2012}). Material and simulation parameters are provided in table 1. The simulation mimics our experiment with 20,000 soft spheres that are dropped into a square container of length 15 cm and a free weight placed on top (see Figure 1b). 

In the experiments, rotational displacement is calculated as the change in the orientation of the cylindrical cavity located within the grain. The angle is then multiplied by the radius of the bead to compare with linear displacements. Measuring only a single axis for the rotations means that we are only capturing 2 of the 3 rotational degrees of freedom: rotations around the axis of the cylindrical cavity remain undetected. In contrast, the simulations allow us to fully track the rotations of the particles. Therefore, when comparing our simulations to the experiments, we calculate the rotational motion using only a single tracked axis (see \href{run:supporting_info.pdf}{supplemental}). 

\begin{figure}[b!]
\includegraphics[width=0.8\columnwidth]{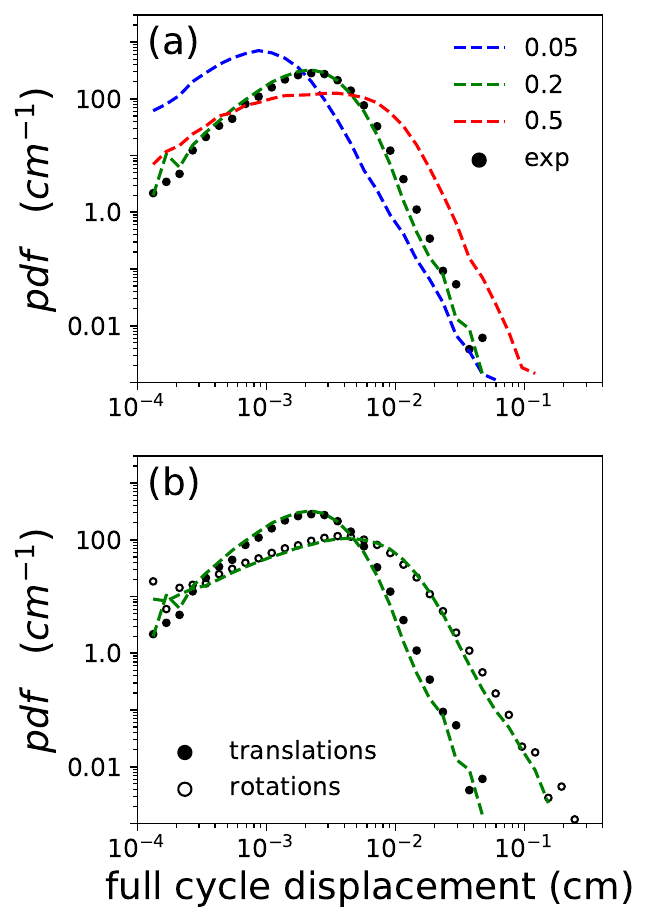}
\caption{\label{fig:two}(a) pdfs of the amplitude of translational displacement after a whole cycle. Colors indicate the static friction coefficient between grains. Black dots correspond to experimental results. (b) Comparison with experimental observations of both the rotational (open circles) and translational (closed circles) displacement. The rotations correspond to the angle between a single-tracked axis in both the experiments and simulations.}
\end{figure}

\section{Results and Discussion}

Verifying the numerical model - Figure 1 shows a projection in the yz plane of each grain’s position in both the experiments (left) and simulations (right) after 500 cycles of compression with an amplitude of A= 0.15 cm (1\% of the container size). The color corresponds to the displacements of the grains when fully compressed. The results show excellent agreement for both the translational and rotational motion between the simulations and the experiments. We see shear zones in the translations at an angle to the compression wall, which are not observed in the rotations. The shear zones indicate that the displacements at maximum compression are highly coupled to each other within each band, whereas the rotations appear to be randomly distributed and uncorrelated with translations. In other work\cite{Benson:2020}, we found the rotations to be correlated with the spatial gradient of linear displacements.

Since rotations are driven by friction between contacts, we expect our simulations to match our experiments at a unique friction coefficient. Figure 2a shows probability density functions (pdfs) of the end-of-cycle translational displacement for 3 values of friction for our simulations. In all cases, the system is compressed at an amplitude of 1\% for 500 cycles, then the pdfs are generated using data from 10 consecutive cycles. We see a large difference in grain displacements as the friction is varied. Specifically, it appears the system systematically gets more reversible as the friction is reduced. That is, the total displacement after a completed cycle is lower for lower friction. By adjusting the friction coefficient, we match our experiments to our simulations nearly perfectly for a static friction coefficient of $\mu_S = 0.2$.  Figure 2b shows the results for $\mu_S = 0.2$ alongside the experiments for both the rotational and translational displacement of the grains. In the experimental setup, the rotations are affected due to the cylindrical cavity hole locking in place with other grains (see \cite{Peshkov:2019} for details). To negate this effect, we restrict our analysis of rotations in experimental data to the grains that do not contact through the hole. The translations are unaffected by the hole contact. Furthermore, Figure 2b shows that the value of friction that best matches the translational displacements predicts the rotational motion as well. We observe a wider distribution for the rotations when compared to the translations, and the grains appear to rotate a larger distance than they translate over an entire cycle. This stems from the irreversibility of rotations compared to the translations\cite{Peshkov:2019}.

As the system is compressed, we expect it to evolve asymptotically towards some unknown steady state as it forms a memory of its input\cite{Bandi:2018}. Accordingly, we present in Figure 3 the mean displacements of the grains, between each cycle, as a function of the cycle number. We subtract the final “steady-state” value, which is the average of the motion for the final 20 cycles, to reveal a power-law evolution, similar to what is done in ref\cite{Bandi:2018}, for both translations and rotations with exponents of $\sim 0.7$ and $\sim 0.6$, respectively. The difference in the initial amount of translations in the simulations indicates that the initial configuration of the experimental system, produced by stirring and deposition of a top weight, is not fully captured in the simulated initial conditions. We note that both experiments and simulations evolve following the same power-law exponent and converge to the same system state (Figure 3).

\begin{figure}
\includegraphics[width=0.8\columnwidth]{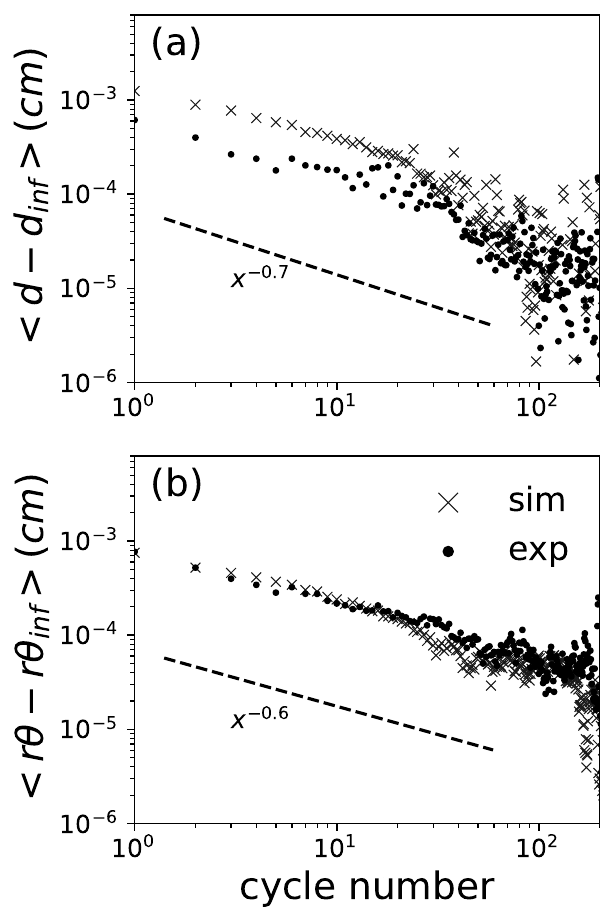}
\caption{\label{fig:three}Mean displacement as a function of the cycle number for the experiments (circles) and the simulations (crosses). The theoretical “steady-state” displacement is subtracted from both curves to reveal a power-law behavior. Dashed lines correspond to the fitted curve, with exponents of approximately 0.7 and 0.6 for the 
translations and rotations, respectively.}
\end{figure}

\begin{figure*}
\includegraphics[width=1.8\columnwidth]{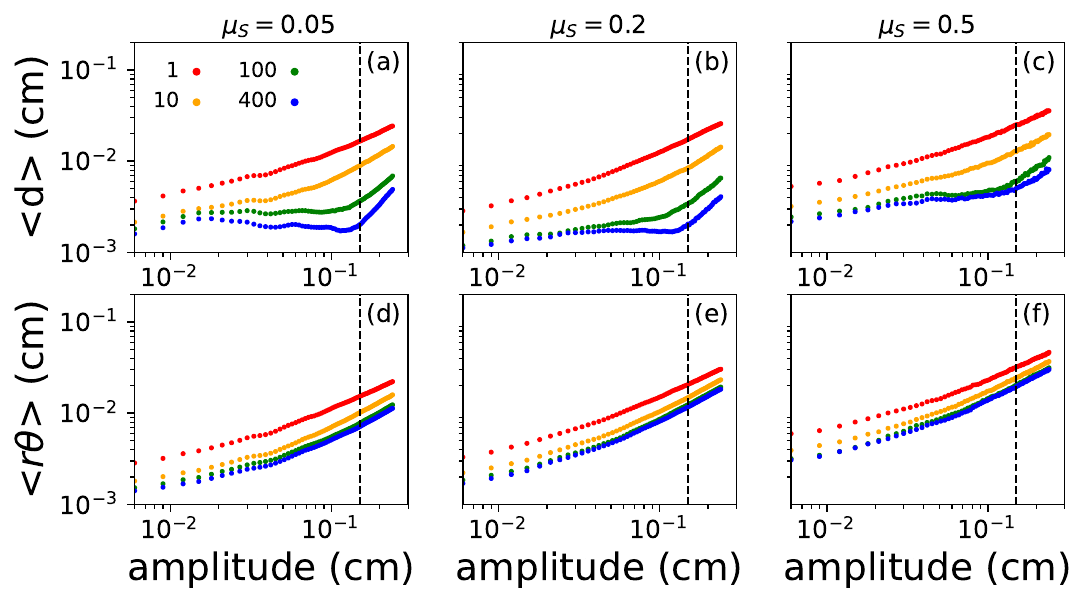}
\caption{\label{fig:four}Parallel memory read-out protocol for translations (a-c) and rotations (d-f). Static friction is varied, with $\mu_S = $ 0.05 (a,d), 0.2 (b,e) and 0.5 (c,f). Colors indicate the amount of training cycles performed before the read-out. The dashed line indicates the amplitude at which the system was trained. A dip at the training amplitude appears for translations. Two distinct power laws emerge in the rotations.}
\end{figure*}

Encoding and reading memory - For the rest of this Letter, we will use the simulations to explore memory formation in a cyclically compressed granular system. Using the simulations allows us to perform a parallel read-out of the memory commonly used in these types of studies\cite{Paulsen:2014}\cite{Adhikari:2018}\cite{Fiocco:2018}. The training protocol takes a trained configuration (i.e. the positions, forces, and orientations) of a system, and conducts a parallel sweep by conducting a full compression cycle $0 \rightarrow A’ \rightarrow 0$, with a variable $A'$. The displacements are calculated by taking the position after each cycle. In our case, we perform an initial training at a compression amplitude $A$ = 1\%~(0.15~cm) and then use that configuration of grains and inter-particle forces repeatedly in a sweep of compressions $A’$~=~0.05\%~to~2.5\%.  Figure 4 shows the read-out results for our simulations as a function of both the friction and the amount of training cycles. We consider motion up to 400 cycles; in this regime, several macroscopic parameters including the packing fraction, and compression force on the wall, remain unchanged from cycle to cycle. Thus, we argue that the system has reached a sufficiently steady-state up to the resolution of our measurements.

Memory of the compression amplitude is apparent in the translations as a dip in the mean displacement of the grains and subsequent increase beyond the 1\% marker. Moreover, we see that the memory formation appears to be stronger the more cycles the system has experienced. However, for the high-friction case, the material reaches a “steady-state” after 100 cycles in which the change in the memory effects is not discernible anymore. The dip at 1\% appears less pronounced for the high-friction case (Figure 4c), suggesting that the motion is not periodic in the same way as for low friction (Figures 4a,b). It is important to note that at low friction, the system presents identical mean displacements for different perturbation amplitudes around the dip. This implies that it is not possible to determine the state of the system from a single perturbation.

Since the rotations also have a similar, power-law evolution in the average displacements as a function of cycle number, we expect the memory read-out to be similar to that of the translations. Figure 4d-f shows the read-out of the mean rotations for static friction of 0.05, 0.2, and 0.5, respectively. Immediately, we observe that the rotations appear not to have a similar memory signature in their displacements. However, there is a distinct power-law behavior that undergoes a slope change as the amplitude is increased that gets more pronounced as the system experiences more training cycles. Specifically, after 400 cycles, we find an exponent of $\sim 0.7$ for small amplitudes and then $\sim 1.4$ for large amplitudes, 2 times the small-amplitude exponent. One might suspect that the change in behavior of rotations is caused by the change in the translational motion. However, it appears that the rotational memory behaves similarly for all friction values, which is not the case for translational displacement. Even with the change in the power law, it appears that the memory of the drive does not seem to be present in the average rotational motion of the grains; however, the evolution or formation of the memory could be present in collective rotations\cite{Pires:2020}\cite{Stager:2016} at longer length scales, which is a topic of future work.

\begin{figure}
\includegraphics[width=\columnwidth]{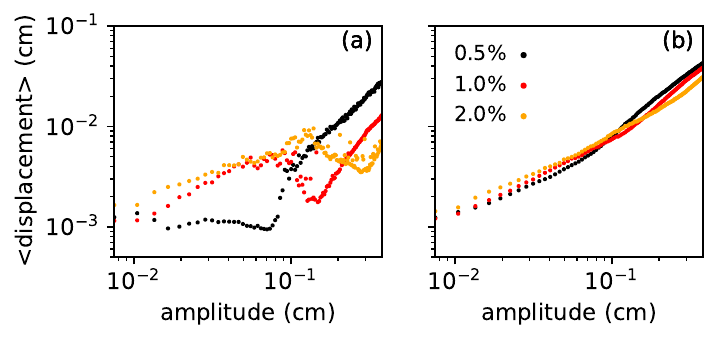}
\caption{\label{fig:5}Parallel memory read-out protocol for translations (a) and rotations (b) for static friction $\mu_S = 0.2$. The colors indicate the amplitude the state was prepared at.}
\end{figure}

Figure 5 shows how the reversibility is affected by different preparation amplitudes. We have 3 systems that were compressed at 0.5\%, 1\% and 2\%, and performed the same parallel readout to quantify the reversibility. We did this for $\mu_S = 0.2$, the friction that best corresponds to the experiments. Clearly, the reversibility relies heavily on the prepared amplitude (hence the dip in the curves). Additionally, it appears that the system gets more irreversible at higher compression amplitudes, and we see a very pronounced minimum for higher amplitudes. This makes sense as the sample gets less reversible for higher amplitudes. However, looking at the rotations, we see that the power-law increase has a similar exponent for amplitudes higher than the prepared amplitudes for all three curves ($\sim 1.4$)

The memory we observe here is unique from previous interpretations such as “return point memory”\cite{Keim:2020}. Instead, this phenomenon appears more like "novelty detection"\cite{Gold:2019}. It appears that the reversible steady-state depends on the entire trajectory ($0 \rightarrow A' \rightarrow 0 $). That is, each grain has a unique trajectory that it will respond to at amplitude A, and any deviation from this amplitude will form a new, different trajectory. This is clear in figure 8, where we plot trajectories within a cycle for three consecutive cycles for all the friction values. We see that the paths overlap most for the lowest friction case. This is also the friction that has the most pronounced dip in Figure 4a. These repeated trajectories go away for higher friction values. Moreover, the area within the trajectories goes down for the highest friction. This could directly correlate with the absence of the dip in the $\mu_S = 0.5$. This phenomenon is similar to, but not equivalent to the “return point memory” interpretation. For the rotations, we do not see any structure in the trajectories (see figure 9). Instead, we see that the higher friction cases $0.2,0.5$ have trajectories that appear uncorrelated with the wall compression.

\section{Conclusion}

We have demonstrated that soft-sphere collisional simulations successfully capture the quasistatic rearrangements and rotations of a jammed granular system.  We have shown that 3D dense frictional grains exhibit a memory effect when subject to boundary-driven periodic forcing. We verified that the rotational displacements do not encode memory in the same way as the translations, which could be due to the lower overall reversibility of rotations compared to translations; however, we do see a difference in mean rotational irreversibility both below and above the training amplitude in the form of a power law. Moreover, we have found that at low and intermediate friction values, the translations appear to be most reversible at the prepared amplitude. Our work further probes the reversibility of granular rotations in jammed materials. Specifically, we emphasize that irreversible motion still has some kind of memory signature embedded in the rotational displacements that is fundamentally different from translations, whose memory can be read out by probing reversibility. Moreover, our observation of memory in a quasi-static system poses the challenge of how to extract this memory from measurements on the static particle configuration alone, e.g. by using machine-learning\cite{Zhong:2020}.

We would like to acknowledge discussions with members of the Losert lab. The authors acknowledge the University of Maryland supercomputing resources (http://hpcc.umd.edu) made available for conducting the research reported in this paper. ZB was supported by the National Science Foundation graduate research fellowship program. WL and AP were supported by National Science Foundation Grant No. DMR-1507964. AP was also supported by DMR-1809318.  

\onecolumngrid
\appendix

\section{Simulation parameters}

The simulation parameters are given in table I. The simulations are performed using a soft-sphere model, where a linear repulsion force is proportional to the amount of overlap between grains, and the torque is a linear spring force proportional to the deviation from the initial point of contact between two grains. In the table, $\Delta t$ and $k_n$ correspond to the time step and normal spring constant, respectively. We use a spring constant and time step such that grains maintain an overlap $<<$ 1\% of their radius. Although in practice, the acrylic beads used in the experiments are orders of magnitude more stiff, we see this as a fair approximation that effectively reproduce the translational and rotational motions observed in the experiment.

\begin{table}[h]
\centering

\begin{tabular}{|c|c|c|c|}
\multicolumn{4}{c}{\large Simulation Parameters} \\ \hline

	$\Delta t$ 		& 	$25\mu s$ 	& 	$\mu_S$ & $0.2$		\\ \hline
	$k_n$			& 	$4e6 g/s^2$	& 	$\mu_R$ & $0.01$ 	\\ \hline
	$C_n$			& 	$0.2$		& 	$\mu_T$ & $0.001$	\\ \hline
	$C_t$			& 	$0.5$		& 	$\beta$	& $0.5$		\\ \hline
\end{tabular}
\caption{The table containing the physical parameters used in the simulations.}
\end{table}

The tangential spring constant is defined as $\frac{5}{7}k_n$. $C_n$ and $C_t$ are the restitution coefficients. Given that our simulations are quasi-static, we do not expect the restitution to play much of a role in the dynamics. The static, rolling and twisting friction values are labeled as $\mu_S$, $\mu_R$, and $\mu_T$, respectively. $\beta$ is the rotation dashpot model’s shape parameter (see ref \cite{Schwartz:2012} for further information on these parameters).

\section{Calculating granular rotations}

\begin{figure}[b]
\includegraphics[width=0.4\columnwidth]{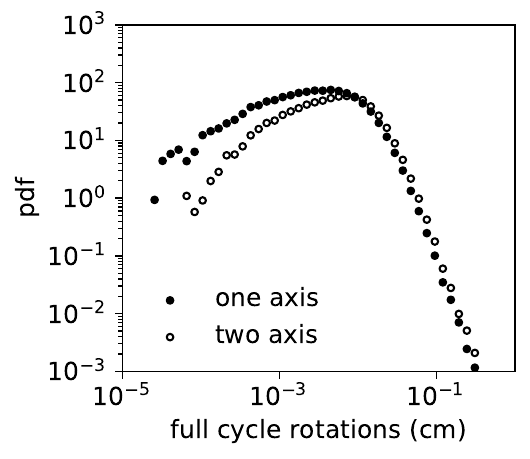}
\caption{\label{fig:5} pdf of the full cycle rotation using one axis (solid circles) or two axes (empty circles) when computing the magnitude.}
\end{figure}

\begin{figure*}[t]
\includegraphics[width=\columnwidth]{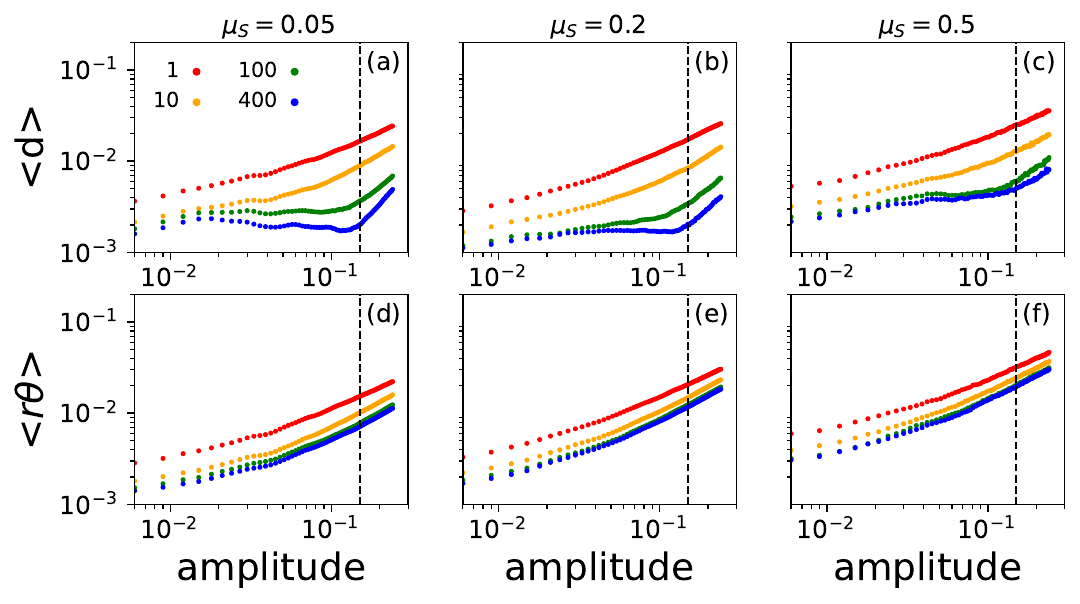}
\caption{\label{fig:one}Memory readout using a symmetric drive ($0\rightarrow A/2 \rightarrow A/2 \rightarrow0$) for different friction values. The color indicates the amount of training cycles performed. The top row is the displacements, and the bottom row is the rotations.}
\end{figure*}

Figure 6 shows a probability distribution function for rotational displacements after a completed cycle using both or a single axis to compute the magnitude from the simulation data. Note that the single axis curve can be effectively rescaled to the two axes calculation by a division of the rotation amplitude, consistent with a measure of Euclidian distances of a single or two random variables.  Rotation about a single axis is given by the following equation: 

\begin{equation}
sin(\theta) = |\hat{a}_1\cdot \hat{a}_2|
\end{equation}

Here, $\hat{a}_1$ and $\hat{a}_2$ are the initial and final orientation of the grain, and $\theta$ is the angle between them. For the simulations, this calculation is done by tracking a single principal axis in time that is rotated based on the torques calculated. Note here that the rotation is limited to 90 degrees due to the symmetry of a rotation about a single axis.

For the rest of paper, the simulated rotation is calculated by a rotation matrix defined as:

\begin{equation}
RP_1 = P_2
\end{equation}

$P_1$ and $P_2$ are now a 3 by 3 matrix containing 3 tracked principal axis of the each of the spheres. The rotation matrix $R$ maps the initial principal axis to the final. The angle is calculated as follows:

\begin{equation}
2 cos(\theta) = Trace(R) - 1
\end{equation}

Here, the rotations are limited by 180 degrees since all principal axes are labeled and tracked throughout the simulation.

\section{Symmetric drive}

In the main text, the shear moves the lateral wall from position $y = 0$, to an amplitude $A$, and then back to $0$. This drive is purely asymmetric. Other work on memory also probes symmetric drives, where the compression moves the wall from ($0\rightarrow A/2 \rightarrow A/2 \rightarrow0$). In figure 7, we show what our memory signature looks like for symmetric drives. The main difference is the absent peak around the trained amplitude. Moreover, the system seems to take a lot longer (more cycles) to encode the compression amplitude when compared to the asymmetric drive in the paper. The rotations also do not appear to have any memory signature for this drive.

\section{Trajectories}

Figure 8 plots trajectories of a single grain for three consecutive cycles, for all the friction values. The grains that were selected are within the shear zone of the material (see Figure 1 for reference). The plot is a projection into the YZ plane for visualization.  There is a clear distinction between the three friction values. For the low friction, the trajectories overlap cycle to cycle with a hysteretic loop. As friction is increased, there are two things to notice, (1) the area within the loop decreases, and (2) the trajectories don’t appear to overlap from cycle to cycle. This difference could be the reason why we see a dip in the memory sweep (Figure 4) for $\mu_S = 0.5$ and $0.2$, whereas it’s absent for $0.5$.

\begin{figure*}
\includegraphics[width=0.8\columnwidth]{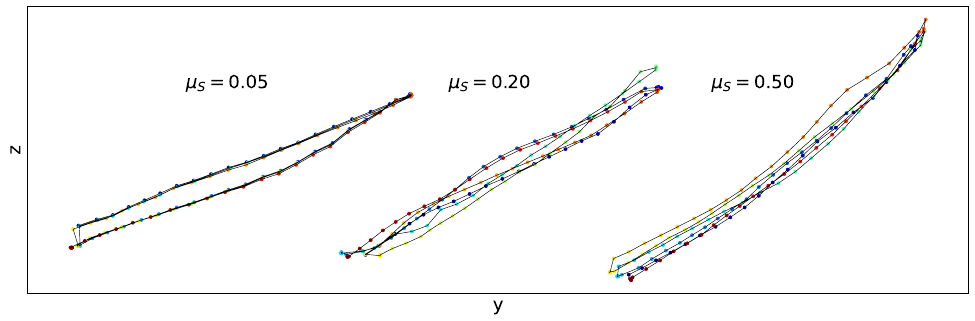}
\caption{\label{fig:one} YZ projection of the average displacement of a small group of grains for 3 consecutive cycles. The color (blue to red) corresponds to the beginning of the first cycle (blue) to the end of the third tracked cycle (red). The curves correspond to the different friction values. The first cycle tracked here begins after 400 completed compression cycles.}
\end{figure*}

\begin{figure*}
\includegraphics[width=0.8\columnwidth]{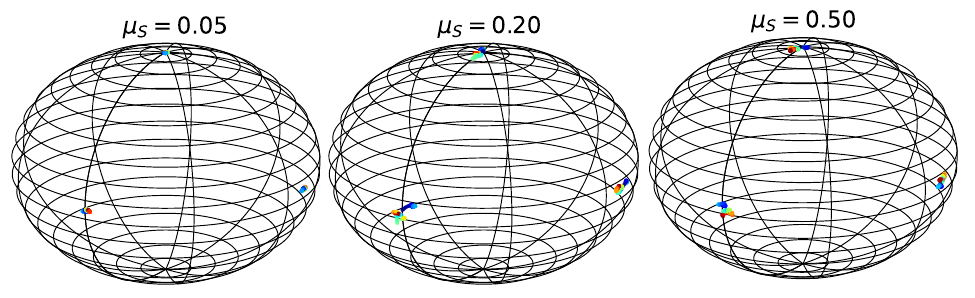}
\caption{\label{fig:one} Orientation trajectory of a grain for three consecutive cycles. The colors (blue to red) correspond the beginning of the first cycle (blue) to the end of the third cycle (red). The 3 spheres are the different friction values. The first cycle tracked here begins after 400 completed compression cycles.}
\end{figure*}

Figure 9 presents the trajectories of the orientations of a single grain within the sample. These grains are also taken within the shear zone of the material. We notice no such structure in the rotations that we see in the translations. Moreover, it appears that the rotations appear diffusive for the higher frictions. That is, they seem to trace out random trajectories. For the low friction, we see almost no rotation for many of the grains. This could be caused by  the low frictional force between the grains.

\bibliographystyle{aipnum4-1}
\bibliography{references_database}

\end{document}